# Metasurfaces for Quantum Photonics

Alexander S. Solntsev[1], Girish S. Agarwal[2], and Yuri S. Kivshar[3]

[1]School of Mathematical and Physical Sciences, University of Technology Sydney, Ultimo NSW 2007, Australia

[2]Texas A&M University, College Station TX 77843, USA

[3]Nonlinear Physics Centre, Australian National University, Canberra ACT 2601, Australia

## Abstract

Rapid progress in the development of metasurfaces allowed to replace bulky optical assemblies with thin nanostructured films, often called *metasurfaces*, opening a broad range of novel and superior applications to the generation, manipulation, and detection of light in classical optics. Recently, these developments started making a headway in *quantum photonics*, where novel opportunities arose for the control of nonclassical nature of light, including photon statistics, quantum state superposition, quantum entanglement, and single-photon detection. In this Perspective, we review recent progress in the field of quantum-photonics applications of metasurfaces, focusing on innovative and promising approaches to create, manipulate, and detect nonclassical light.

**Introduction**

The recent years showcased a resurgence of diffractive optics, enabled by advances in nanofabrication of large-area arrays of metallic and dielectric nanoresonators with high precision, reasonable throughput, and relative ease of production. These developments ushered a new area of the so-called *flat optics*, with the key components called *metasurfaces* being two-dimensional structures composed of optically thin arrays of scatterers, such as subwavelength-sized antennas, and they are increasingly used to replace whole sets of traditional optical elements [Yu 2014, Chen2016, Kruk2017, Li_G2017, Chang2018, Rubin 2019, Osborne2019]. These devices enable efficient beam steering, local control of optical polarization, and enhancement of emission and detection of light [Chen2018, Kamali2018, Overvig2019, Kang2019].

Metasurfaces possess unique capabilities of fully controlling light within a subwavelength layer [Luo2018]. That includes wavelength- and polarisation-selective control of complex diffraction. Moreover, metasurfaces enable new physics and a range of phenomena that are distinctly different from what can be achieved in bulk optics or three-dimensional metamaterials. One such example is the generalized law of reflection and refraction, where metasurfaces can be utilised for the redirection of an incident beam by employing antenna arrays with prescribed phase gradients, while ensuring unprecedented design flexibility with complete control of both amplitude and phase. Metasurfaces can also tailor near-field response, which is crucial when dealing with optical sources and detectors, enabling perfect absorption, emission enhancement, and detailed design of light-matter interaction properties. Note that the metasurfaces use dielectric nanoscale elements [Kruk2017] and thus response is sensitive to the geometry rather than wavelength and therefore useful over a very wide spectral range unlike plasmonic materials whose utility is restricted by the localized plasmonic wavelengths.

Metasurfaces have now become a staple in classical optics, and now there is increasing interest in bringing novel functionalities enabled by flat photonics to the realm of *quantum optics* [Li_C2020]. The quantum optical technologies require sources of single photons, entangled photons, and other types of *nonclassical light*. These also require newer methods of detection. The quantum states could be based on different degrees of freedom of light polarization, direction, orbital angular momentum. For the realization of each of these, metasurfaces have great potential as we discuss below in this Perspective. We first notice that the demonstration of the quantum interference [Hong 1987; Pan2012] of two independent photons at a classical optical device – a beam splitter, and the generation of entangled states became a milestone in the field of quantum optics. A beam splitter is a simple device where one can only change its reflectivity and thus does not have much functionality. Metasurfaces have much broader functionality and have great potential to manipulate single photons and produce a wide variety of multiphoton entangled states. For example, the metasurface can entangle the orbital and spin degrees of freedom, whereas the beam splitter cannot. Some of these applications have started appearing in last few years or even months, and they would be discussed below. We notice further that the detection of nonclassical light carrying the useful information like small phase changes introduced by an object requires special methods as the measurement of intensity does not give complete information on nonclassicality. The most prominent nonclassical detection methods are the measurements of the intensity correlations and statistics [Varro2008] and the homodyne detection analysis [Collett1987]. Ideally, one would like to have full photon statistics, and with the availability of the single-photon edge sensors, it is becoming possible to distinguish between the signals say with one photon and two photons [Magaña-Loaiza2019]. Other types of quantum measurements, like that of the weak values associated with the state [Aharonov1988], are being successfully implemented using metasurfaces.

Nonclassical optics has a wide range of applications in quantum communications [Gisin2002, Gisin2007], computation [Slussarenko2019] and sensing [Giovannetti2004, Hudelist2014, Solntsev2018]. Utilising quantum-mechanical effects such as photon indivisibility, quantum superposition and entanglement, offers a vision of a quantum revolution, with qualitative improvements in a wide range of technologies. Quantum-optical internet could provide security ensured by the laws of physics, with small-scale implementations already showing significant progress [Yin2017, Yin2020]. Distributed quantum computation facilitated by quantum optical links is poised to enable ultrafast simulations of complex quantum systems with follow-up advances in medicine, chemistry, and material science [Ladd2010]. Quantum-enhanced optical sensing is already revolutionizing cutting edge measurements [Samantaray2017, Anderson2017, Dowran2018]. All of the above can benefit significantly from the physics of metasurfaces providing compact, fast and precise control of quantum photonic states.

In this Perspective, we review major directions in this recently emerged research field, by looking at how metasurfaces can be employed to generate, manipulate and detects nonclassical light. We discuss *three major directions* illustrated schematically in **Fig. 1**. The generation includes the integration of metasurfaces with single-photon emitters based on quantum dots [Shields2007] and solid-state colour centres [Mizuochi2012, Tran2016], two-dimensional arrangements of subwavelength nonlinear sources of nonclassical light based on spontaneous parametric down-conversion [Marino2019], and implementations of metasurface concepts in atomic quantum optics [Zhu2020]. The manipulation of photonic states focuses on the utilization of metasurfaces for the control of quantum interference [Beugnon2006, Wen2013] and quantum entanglement [Moreau2019, Moreau2019_2], enabling a range of application such as quantum information processing and compete quantum state tomography. Metasurfaces can be utilized for quantum-enhanced sensing [Gefen2019], weak measurements that do not destroy quantum states [Ritchie1991, Pryde2005, Salvail2013], perfect absorption [Baranov2017] of single photons and quantum imaging [Brida2010].

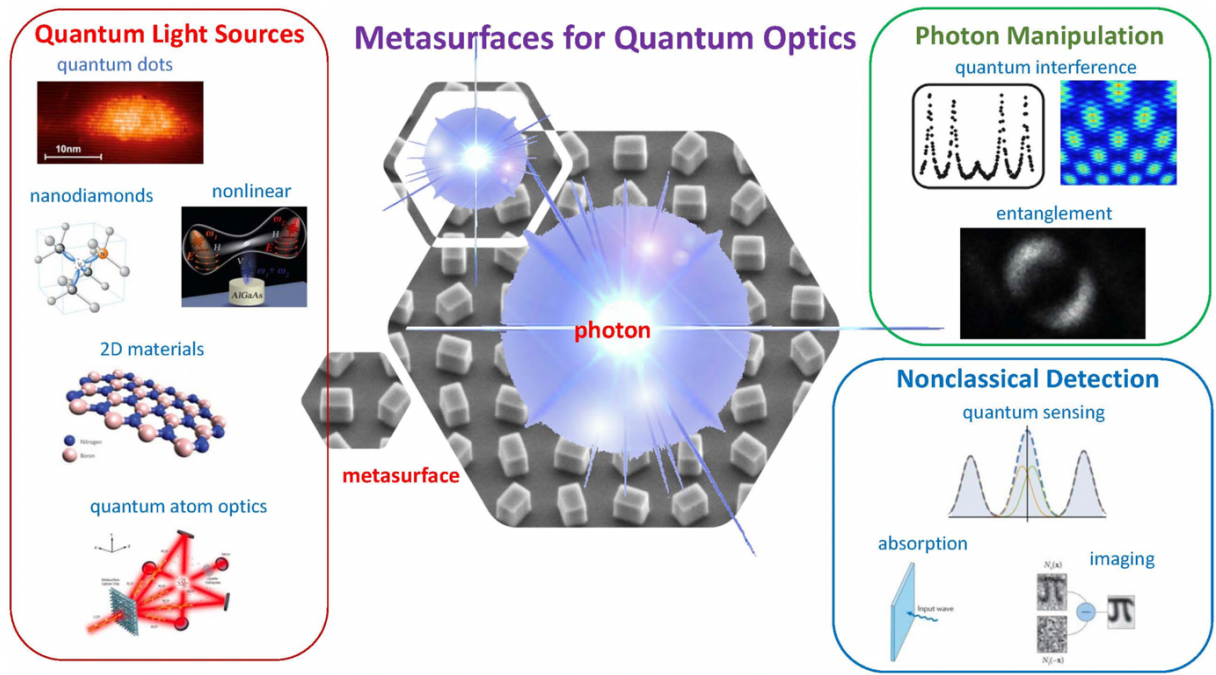

**Figure 1 | Quantum optics with metasurfaces.** Illustration of different cases of interaction between optical metasurfaces and photons, and three major directions discussed in this Perspective: quantum light sources, photon manipulation, and nonclassical detection of light. Insets in the figure adapted from Refs. [Shields2007, Mizuochi2012, Marino2019, Tran2016, Zhu2020, Beugnon2006, Wen2013, Moreau2019, Gefen2019, Baranov2017, Brida2010].

**Quantum Light Sources**

Nanophotonics is attractive for enhancing the generation of single photons and photon pairs for further processing and coupling to optical circuitry, and it is considered as one of the critical enabling technologies for quantum communication and computation systems [Pelton2015]. In this, context, the coupling of nonclassical light sources to metasurfaces enables enhanced emission control beyond what can be provided by single nanoparticles. The use of metasurfaces in nonlinear and atomic quantum light systems significantly increases their nonclassical functionalities.

***Single-photon emitters.*** The conventional approach to the realization of a single-photon source is to make use of spontaneous emission from a single two-level system emitting one photon at a time – the so-called *quantum emitter* – that can be selected from various structures, including dye molecules, quantum dots, and colour centres in crystals. The advantage of such emitters is that one has a source of single photons with a well-defined wavelength, and these are deterministic sources of single photons. Besides a number of fundamental aspects such as cooperative and many body effects can be investigated which can lead to both emitter -photon and emitter entanglement. However, the radiative lifetimes of quantum emitters, often being of the order of 10 ns, are too long to meet the speed requirements of optical communication and information processing systems. The rate of spontaneous emission can, however, be increased by placing a quantum emitter in a suitable photonic environment with an increased electromagnetic local density of states, as discussed in literature [Agarwal1975, Lunnemann2016]. This concept has been reviewed by Vaskin et al. in the context of light-emitting metasurfaces [Vaskin2019]. In this approach shown schematically in **Fig. 2a**, a metasurface can be employed as "special photonic environment," which provides novel ways to manipulate and control quantum light with flat optics.

Iwanaga et al. [Iwanaga2018] has demonstrated that sparsely distributed quantum dots (QDs) embedded in semiconductors coupled to plasmonic metasurfaces (see **Fig. 2b**), can experience significant increase in the activity of photoluminescence (PL) response, including becoming superlinear with respect to the excitation laser intensity under weak excitation. PL response was examined [Iwanaga2018] at room and cryogenic (9 K) temperatures showing that hot electrons mostly contribute to superlinear PL responses at room temperature, whereas induced transitions between the excitonic levels in the QDs are significant at 9 K. This demonstration may have important implications for the realisation of efficient single-photon-emitting devices.

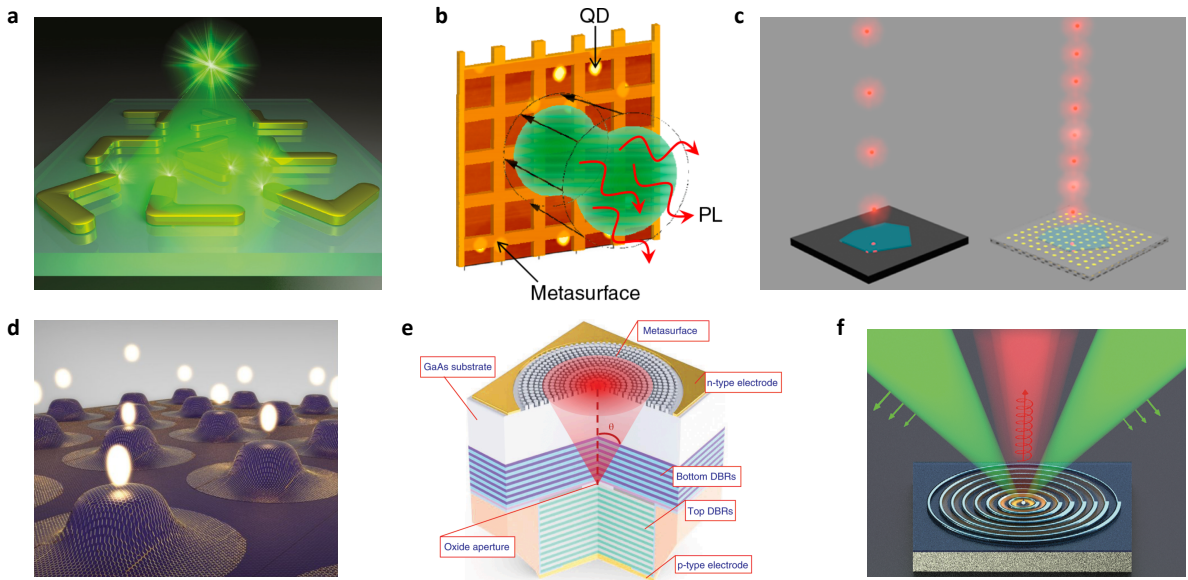

**Figure 2 | Integration of single-photon emitters with metasurfaces. a**, Illustration of a single quantum emitter interacting with metasurface scatterers featuring Purcell enhancement. **b**, Plasmon metasurface coupled to quantum dots for superlinear photoluminescence. **c**, Deterministic coupling of single quantum emitters in 2D hBN to a plasmonic nanocavity array. **d**, Single-photon emitters in 2D hBN activated by a metasurface comprised of silica pillars. **e**, Metasurface for programmable directional emission integrated with vertical cavity surface-emitting laser. **f**, Metasurface-enabled generation of circularly polarized single photons. Figure adapted from: **a**, ref. [Vaskin2019]; **b**, ref. [Iwanaga2018]; **c**, ref. [Tran2017]; **d**, ref. [Proscia2018]; **e**, ref. [Xie2020]; **f**, ref. [Kan2020].

Two-dimensional materials such as graphene, hexagonal boron nitride (hBN), and transition metal dichalcogenides can also serve as single-photon sources. Compared to semiconductor quantum dots, these materials can be integrated more easily with photonic metasurfaces. One of the most straightforward ways to utilise such integration is for Purcell enhancement. Tran et al. [Tran2017] have shown that quantum emitters in two-dimensional hexagonal boron nitride can be deterministically coupled to a plasmonic nanocavity array. As shown in **Fig. 2c**, Purcell enhancement in the weak coupling regime when coupled to a plasmonic metasurface, leads to enhanced emission rates and reduced fluorescence lifetimes. Importantly, in this case, the single-photon statistics is largely preserved.

Following the demonstration in $WS_2$ at a cryogenic temperature by Palacios-Berraquero et al. [Berraquero2017], Proscia et al. [Proscia2018] has shown that single-photon emission from defects in hBN at room temperature can not only be amplified via the integration with a metasurface, but also that the single-photon emitting defects themselves can be induced via strain, when hBN is placed on top of a specially designed metasurface. **Figure 2d** shows an illustration of emitters in hBN activated when strain is introduced by an array of silicon rods. This finding is significant because it allows near-deterministic activation of sites for single-photon emission. Through the combined control of strain and external electrostatic potentials, Proscia et al. [Proscia2018] has demonstrated the realisation of arrays of room-temperature single-photon sources with well-defined positions.

Another important contribution to metasurface optics has recently been revealed by Xie et al. [Xie2020]. As shown in **Fig. 2e**, a metasurface integrated at the back side of the substrate can be combined with a light source for the purpose of beam-shaping. In this design, centrosymmetric GaAs nanopillars of different diameters are used as polarization insensitive meta-atoms. The light source in question is a vertical cavity surface-emitting laser, however the same approach is also expected to work for obtaining the directional emission of nonclassical single-photon sources. This work shows that such metasurface integration enables highly efficient arbitrary control of the emission beam profiles, including self-collimation, and the formation of Bessel and Vortex beams.

Extending this approach to nonclassical light, Kan et al. [Kan2020] has shown that dielectric metasurfaces can be used for the generation of highly directional circularly polarised single photons. As shown in **Fig. 2f**, in this work a nanodiamond containing a single nitrogen vacancy (NV) centre that can emit single photons is placed in the centre of an optical metasurface composed of concentric periodic width-varying dielectric nanoridges atop a thin dielectric film on a metallic substrate. The arrow with a helix in the red beam illustrates a collimated stream of circularly polarised single photons, while the green cones represent a tightly focused radially polarized pump beam. In this configuration the single-photon chirality of 0.8 and high directionality leading to collection efficiency of 92% can be achieved.

***Nonlinear metasurfaces.*** With metasurfaces composed of planar lattices of optical subwavelength resonators, one can miniaturise substantially nonlinear photon sources [Caspani2017, Solntsev2017] that utilize spontaneous four-wave missing (SFWM) or spontaneous parametric down-conversion (SPDC). It might be noted that the SPDC is at the present the most versatile source of heralded single photons and entangled photon pairs over a large region of spectrum; it continues to be the workhorse of the community. Thus, it is desirable to enhance the efficiency of conversion. The use of ultra-thin metasurfaces may open the potential for quantum entanglement between photons at ultra-short timescales across the visible and infrared regions, leading to new opportunities for quantum spectroscopy, sensing, and imaging. This agenda is not yet realized, although the first steps have been taken recently.

Phase-matching-free SFWM has been reported earlier, but only in the degenerate regime [Suchowski2013]. **Figure 3a** illustrates a more recent observation of spontaneous parametric down-conversion free of phase matching with a frequency spectrum an order of magnitude broader than that of phase-matched SPDC [Okoth2019]. By replacing a thin nonlinear membrane with a structured nonlinear metasurface incorporating, for example, AlGaAs nanodisks would enable enhanced photon-pair generation via SPDC [Marino2019]. This will open the potential for generating photons with tailored quantum entanglement for nonlinear quantum spectroscopy, quantum sensing, and ghost imaging [Pittman1995, Strekalov1995, Abouraddy2004].

The generation of photon pairs in nonlinear materials enables the creation of nonclassical entangled photon states. By integrating a metasurface lens with a nonlinear barium borate crystal, as shown in **Fig. 3b,** one can realize a multi-path spontaneous parametric down-conversion photon-pair source. This is promising for high-dimensional entanglement and multi-photon state generation. Such a structure was realized by Li et al. [Li_L2020] for a 10x10 metalens array. They demonstrated four-photon and six-photon generation with high indistinguishability of photons generated from different metalenses. This metasurface-based quantum photon source is compact, stable, and controllable, representing a new promising platform for integrated quantum photonic devices.

The generation and modulation of photonic entanglement was studied by Ming et al. [Ming2020] based on the parametric down conversion processes in a nonlinear plasmonic metasurface. Through flexible nanostructure design, it is possible to tailor the nonlinear photonic interaction in the metamaterial system, and the spatial properties of the generated photonic state can be steered as desired (see **Fig. 3c**). In particular it is possible to give a vortex structure to the second order nonlinearity which facilitates the generation of a range of entangled states of orbital angular momentum. This theoretical framework is based on the nonlinear Huygens–Fresnel principle, and a differential approach utilized to mitigate the intrinsic loss of the system. This platform could be valuable for applications of quantum information processing.

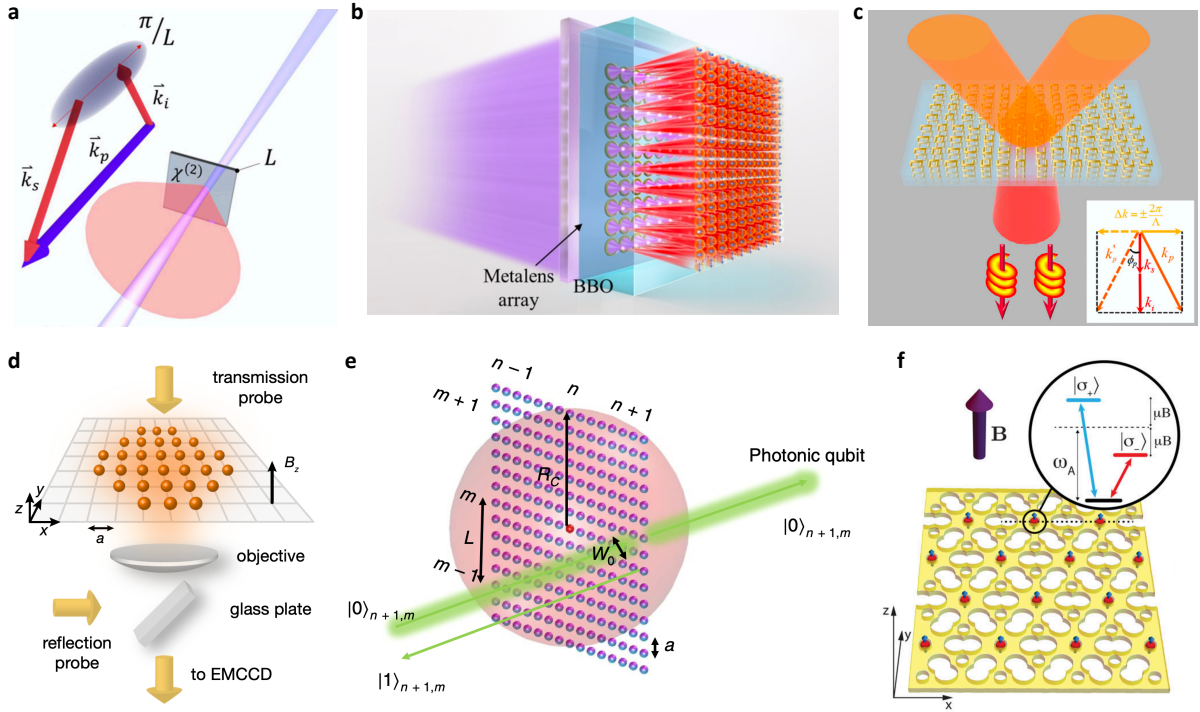

**Figure 3 | Nonlinear and quantum metasurfaces. a**, Generation of entangled photons without momentum conservation. **b**, Metalens multi-photon quantum source. **c**, Illustration of parametric down conversion process in the nonlinear metamaterial. Insert shows the geometry of photon interaction. **d**, The study of the cooperative response of the atomic lattice as a mirror probed by circularly polarized light. **e**, Quantum metasurface based on a two-dimensional lattice of atoms. **f**, Photonic membrane dielectric metasurface with an embedded triangular lattice of quantum emitters. Figure adapted from: **a**, ref. [Okoth2019]; **b**, ref. [Li_L2020]; **c**, ref. [Ming2020]; **d**, ref. [Rui2020]; **e**, ref. [Bekenstein2020]; **f**, ref. [Perczel2020]

***Quantum metasurfaces and atomic arrays.*** The metasurface platform for on-chip quantum state engineering offers a promising route for scaling from two-qubit to many-body entanglement by introducing a multifunctional metasurface. The metasurface approach is especially promising when applied to atomic arrays [Barredo2016, Ballantine2020, Parmee2020]. As these metasurfaces enable photonic multitasking, optical trapping of the qubits can be assigned as additional independent functionality to the primary task of manipulating the flow of photons. These functionalities can be combined with topological properties.

Rui et al. [Rui2020] reported the direct observation of the cooperative and directional subradiant response of a two-dimensional square array of atoms in an optical lattice, a quantum metasurface. They observed a spectral narrowing of the collective atomic response well below the quantum-limited decay of individual atoms into free space. Through spatially resolved spectroscopic measurements, they demonstrated that the array acts as an efficient mirror formed by only a single monolayer of a few hundred atoms. **Figure 3d** illustrates the atomic lattice probed by circularly polarized light. For transmission, both the probing and the residual transmitted fields at the plane of the atoms are collected and imaged. The metasurface contains around 200 atoms in this experiment.

By now, significant efforts have been focused on structuring the properties of light with metasurfaces. Next step is to explore the possibility of generating atom-photon entanglement between atomic metasurfaces and nonclassical light for controlling the many-body entangled photonic states. Such 'quantum metasurfaces' can be realized by preparing and manipulating entangled states of atomic reflectors and scattering light from them, constituting a new platform for manipulating both classical and quantum properties of light. **Figure 3e** illustrates a quantum metasurface realized by entangling the macroscopic response of atomically thin arrays to light [Bekenstein2020]. Such a system allows for parallel quantum operations between atoms and photons as well as for the generation of highly entangled photonic states and three-dimensional cluster states suitable for quantum information processing.

Perczel et al. [Perczel2020] suggested an experimentally feasible nanophotonic platform for exploring many-body physics in topological quantum optics. Their system is composed of a two-dimensional lattice of nonlinear quantum emitters with optical transitions embedded in a membrane metasurface, as shown in **Fig. 3f**. The emitters interact through the guided modes of the metasurface, and a uniform magnetic field gives rise to wide topological band gaps, robust edge states, and a nearly flat band with a nonzero Chern number. In the schematic of **Fig. 3f**, out-of-plane magnetic field B splits the |σ+⟩ and |σ–⟩ atomic transitions.

These results demonstrate efficient optical metasurface engineering based on structured ensembles of atoms and pave the way towards the controlled many-body physics with light, as well as novel light-matter interfaces at the single quantum level.

**Photon Manipulation with Metasurfaces**

One of the earliest demonstrations of the interaction between metasurfaces and nonclassical light is in the field of plasmonics. Altewischer et al. [Altewischer2002] has investigated the effects of nanostructured metallic films on the properties of entangled photons. As shown in **Fig. 4a**, an optically thick metal film perforated with a periodic array of subwavelength holes was placed in the confocal telescope (TEL) in the paths (A1 and A2) of the two entangled photons generated by a barium borate (BBO) crystal and controlled by a half-wave plate (HWP). Such metasurface makes a conversion of photons into surface-plasmon waves, and the question was whether photon entanglement, which is critical for many quantum optical applications, would survive the conversion into plasmons. The result is that the entanglement after the going through the plasmonic metasurface is well maintained, despite substantial losses. This work has opened the field of metasurfaces for quantum optical manipulation, enabling the use of plasmonic and later dielectric metasurfaces in quantum photonics.

*Quantum interference.* The principal manifestations of quantum light are associated with nonclassical interference, which is an enabling phenomenon for the manipulation of quantum states in a variety of applications. Following the prediction of quantum interference in anisotropic environment [Agarwal2000], Jha et al. [Jha2015] has demonstrated theoretically that metasurfaces can be used for remote quantum interference engineering. In this model (see **Fig. 4b**), a specially designed metasurface creates strongly-anisotropic quantum vacuum in the vicinity of a quantum emitter at a distance $d$ much larger than wavelength $\lambda$. The inset shows a three-level atom, with coupled orthogonal transitions whose coupling strength $\kappa$ depends on the anisotropy of the quantum vacuum. In this case, the metasurface can induce quantum interference among radiative decay channels, thus opening the path to the engineering of long-range interactions in solid state systems and quantum atom optics.

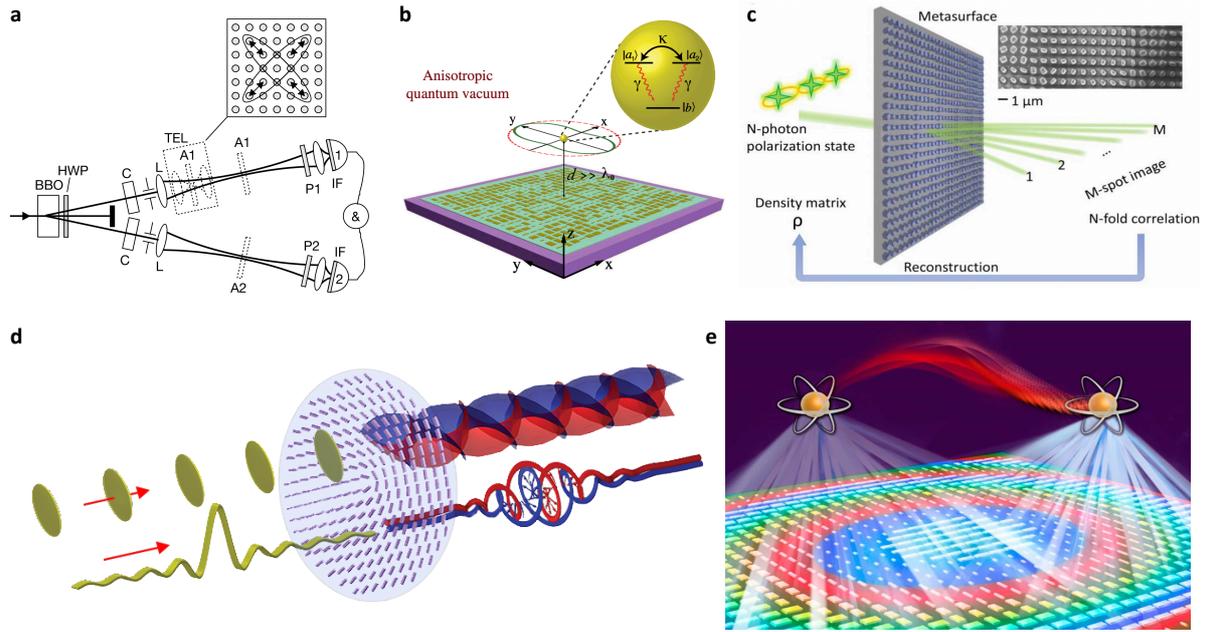

**Figure 4 | Photon manipulation by metasurfaces. a**, Photon entanglement transmitted thorough a plasmonic metasurface. **b,** Metasurface-Enabled long distance quantum interference. **c**, Quantum optical state reconstruction based on a metasurface. **d**, Quantum entanglement of photon orbital angular momentum and spin using metasurfaces. **e**, Quantum entanglement between atomic qubits mediated by a metasurface. Figure adapted from: **a**, ref. [Altewischer2002]; **b**, ref. [Jha2015]; **c**, ref. [Wang2018]; **d**, ref. [Stav2018]; **e**, ref. [Jha2018].

Quantum emitters can also be integrated with anisotropic metasurfaces, which can couple the levels of a quantum emitter through the quantum interference effect leading to remarkable chiral effects. Kornovan et al. [Kornovan2019] has recently predicted that a combination of the metasurface anisotropy and tilt of the emitter quantization axis with respect to the anisotropic metasurface normal results in nonsymmetric dynamics between the transitions from the left-circular state to the right-circular state. It was shown that for four-level atom with an s → p transition placed near anisotropic metasurface, an effective optical activity can emerge due to the anisotropy of the system through the quantum interference of the multiple decay channels of the emitter. This effect opens new avenues for the engineering of nanoscale quantum optical systems.

Further quantum emitter long-lifetime coherence can also be controlled by metasurfaces, as predicted recently. Lassalle et al [Lassalle2020] designed a metasurface to act as a spherical mirror while inverting the absolute rotation direction of the electric field with respect to that of the incident circularly polarized one. This inversion of the electric-field rotation can be achieved by using nanoantennas which act as half-wave plates, as the result of π phase shift between the long and short axes of the nanoantennas. This leads to the acquisition of so-called geometric phase. The result is the creation of a long-lifetime coherence between the two ground states of a quantum emitter, which can be viewed as a first step towards controlling the interactions between several quantum emitters with metasurfaces and generating quantum entanglement.

It has been known for decades that any discrete unitary operator could be realized using conventional optics [Reck1994], however this approach would remain difficult to scale before the arrival of nanophotonics. Recent advances in nanotechnology enabled the integration of beam-splitters and couplers in tailored plasmonic structures, yet such miniaturization came at the cost of material losses and complex photon-plasmon coupling interfaces that restrict the platform scalability [Xu2018]. Wang et al. [Wang2018] revealed an opportunity to break away from conventions of lengthy sequential implementations or lossy plasmonic designs and realized several multiphoton interferences in a single flat all-dielectric metasurface. This scalable approach is based on parallel quantum state transformations encoded in multiple metagratings across the photon beams, taking advantage of the transverse spatial coherence of the photon wavefunctions extending across the beam cross-section. This principle required a nontrivial development for the application to multi-photon states due to a high dimensionality of Hilbert space spanned by the photon number and nonclassical multi-photon interference features. This same approach can also be extended to arbitrary polarisation state manipulation [Lung2020].

Metasurfaces should allow the reconstruction of the total multi-photon quantum state, including phase, coherence, and multi-particle entanglement. Wang et al. [Wang2018] realized an all-dielectric metasurface that spatially splits a tomographically complete set of components of a multi-photon polarization state, such that a simple averaging measurement of correlations with polarization-insensitive on-off detectors enables the accurate reconstruction of a multi-photon density matrix. **Figure 4c** shows a metasurface imaging multi-photon quantum-polarization states, where an input N-photon state is encoded in polarization. The correlation measurements between M output spots enable full reconstruction of the input N-photon quantum density matrix. Top-right inset shows a scanning electron microscope image of the fabricated all-dielectric metasurface.

***Quantum entanglement.*** Metasurfaces have recently had a number of realisations for the manipulation of quantum entanglement, a key resource in quantum optics. This is because specially designed metasurfaces enable the manipulation of the spin-orbit interaction which earlier was done using q-plates [Nagali2009]. **Figure 4d** shows the recent experiments on the generation of entanglement between the spin and orbital angular momenta of photons by Stav et al. [Stav2018]. In this work, the photons from a pair were split – directing one through a unique metasurface and the other directly to a detector to signal the arrival of the other photon. Then the photon that passed through the metasurface was measured to find that it acquired orbital angular momentum and that it became entangled with its spin. In the second experiment, the photon pairs passed through the metasurface and were measured using two detectors to show that they had become entangled: the spin of one photon became correlated with the orbital angular momentum of the other photon, and vice versa.

It has been established [Biehs2017, Jha2018] that a metasurface can be employed to mediate quantum entanglement between two qubits trapped on a metasurface and separated by macroscopic distances, by engineering their coherent and dissipative interactions. As an example, Jha et al [Jha2018] modelled two far-away trapped atomic qubits as positioned at a macroscopic distance from the metasurface. The metasurface shown in **Fig. 4e** is designed such that the spontaneous emission from the source qubit is efficiently directed toward the target qubit at the single-photon level. As a result of this interaction, quantum entanglement between the two qubits emerges instantly and lasts much longer than the lifetime of individual qubits. As such, spatially scalable interaction channels offered by the metasurface enable the robust generation of entanglement.

These examples demonstrate that metasurfaces can offer significant advantages for nonclassical manipulation of light, including the intricate control of quantum correlations and entanglement.

**Nonclassical Photon State Detection and Excitation with Quantum Light**

In classical optics, metasurfaces show extraordinary abilities to manipulate the phase and polarization of light, and their applications can be extended to the quantum optics, including nonclassical detection. In this context, metasurfaces improve a range of techniques, including weak measurements, interferometry-based sensing, quantum absorption and quantum imaging.

***Quantum sensing.*** One of the important concepts in quantum optical detection that can be enhanced by metasurfaces is weak measurements. Chen et al. [Chen2017] employed dielectric metasurface with a tiny phase gradient that keeps the measured system almost undisturbed, and thus may simplify existing schemes in quantum weak measurements. In quantum weak measurements, three stages are generally involved: First, a measured system is prepared in the initial state, then a weak coupling of observable is introduced by a detector, and finally, the system is selected as a final state. In the procedure of measurement, it is a critical need to introduce a weak coupling and keep the measured system almost undisturbed. Here, the desired coupling strength can be obtained by tailoring the shape and size of the structural units of the dielectric metasurface.

**Figure 5a** shows the metasurface version [Chen2017] of the weak measurement experiment [Aharonov1988, Ritchie1991, Pryde2005, Salvail2013]. The initial state of the photon is preselected by the Glan Laser polarizer (GLP1). The dielectric metasurface (MS) generates a small space-variant phase and plays the role of the weak magnetic field. The final state is post-selected by the Glan Laser polarizer (GLP2). Here, the dielectric metasurface introduces tiny momentum shifts to the photons. By designing the structure of the metasurface, any desired weak coupling strength between the device and the system can be obtained. In general, the tiny momentum shifts are introduced by different interferometer systems in quantum weak measurements. The weak measurements are especially useful for resolving two nearby quantum states on the Poincare sphere.

Next, we mention the recent demonstration of a hybrid integrated quantum photonic system with potential applications in quantum sensing. This metasurface is capable of entangling and disentangling two-photon spin states at a dielectric metasurface, as shown in **Figs. 5b,c**. There, a path-entangled two-photon N00N state with circular polarization that exhibits a quantum Hong–Ou–Mandel interference visibility of $86 \pm 4\%$ is generated via the interference of single-photon pairs at a nanostructured dielectric metasurface. Georgi et al. [Georgi2019] demonstrated nonclassicality and phase sensitivity in a metasurface-based interferometer with a fringe visibility of 86.8 ± 1.1% in the coincidence counts. Such a high visibility proves the metasurface-induced path entanglement inside the interferometer. This approach offers exciting applications in sensing and quantum measurements based on interferometry. For example, a phase changing object on the metasurface would introduce a relative phase between the two components of the output state in **Fig 5a** resulting in a sensitive phase sensor. Another interesting possibility is to have an interferometer with two metasurfaces with the phase object in between. The monitoring of changes in the output state would enable highly sensitive phase measurements up to the Heisenberg limit [Leibfried2004]. The generated entangled states could also be exploited in entangled microscopy to improve resolution [Ono2013].

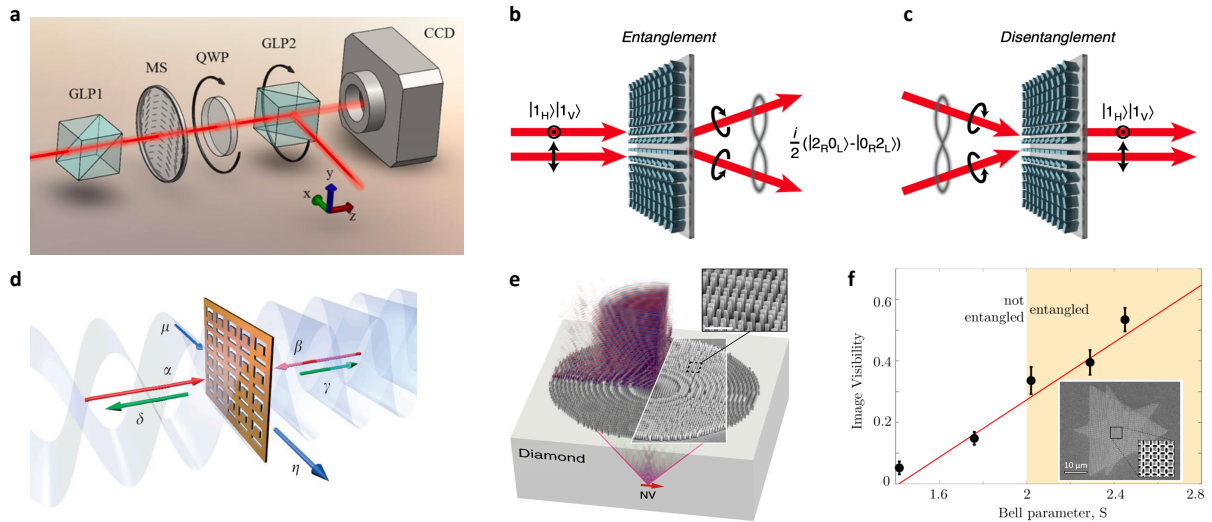

**Figure 5 | Detection of nonclassical light with dielectric metasurfaces. a**, The photonic metasurface version of the experiment on weak measurements. **b,c**, Spatial entanglement and disentanglement of a two-photon state at a metasurface. **d**, Schematic of metamaterial input and output ports. Interaction of two coherent beams in a thin absorber, here represented as a plasmonic metamaterial: the film is described as a lossy beam splitter with two input photon channels, two photon output channels and plasmon input and output channels **e**, Diamond metalens to collect and collimate the emission of an individual NV centre **f**, Imaging with entangled photons: Image visibility for the triangle image plotted vs. the herald photon polarizer angle for increasing degree of entanglement as measured by the Bell parameter *S*. Insert shows the metasurface used in the experiments. Figure adapted from: a, ref. 21 [Chen2017]; b,c, ref. [Georgi2019]; d, ref. [Roger2015]; e, ref. [Lyons2019]; f, ref. [Huang2019].

***Photon absorption.*** Many concepts in the physics of metasurfaces depend on the ability to absorb light very efficiently, and metasurfaces provide novel approaches such as coherent absorption. Extending the control of absorption down to very low light levels and eventually to the single-photon regime is of great interest.

The coherent photon absorption with quantum light was first suggested by Huang et al. [Huang2014]. Later, Roger et al. [Roger2015] demonstrated the coherent absorption of single photons in a deeply subwavelength 50% absorber and revealed that, while the absorption of photons from a travelling wave is probabilistic, standing wave absorption can be observed deterministically, with nearly unitary probability of coupling a photon into a mode of the material, see **Fig. 5d**. These results bring a better understanding of the coherent absorption process, which is of central importance for light harvesting, detection, sensing and photonic data processing applications.

A more recent work by Lyons et al. [Lyons2019] has extended this concept to photon pairs. Here, a metasurface is utilised to coherently absorb two-photon states with 40% efficiency. This concept is very promising, since multiphoton absorption processes have a nonlinear dependence on the number of photons, and engineering devices capable of absorbing pairs of photons efficiently is very challenging. In this paper, the demonstration of coherent absorption of N = 2 N00N states makes it possible to enhance the number of two-photon states that are absorbed by up to a factor of 2 with respect to a linear absorption process. This result is promising for applications where multi-photon absorption is important but limited by damage induced by high peak powers traditionally used in multiphoton experiments. These experiments lead to the possibility of manipulating quantum states like squeezed states by using the coherent perfect absorption and metasurfaces.

***Quantum imaging.*** The process of making a visual representation plays an important role in optics, and metasurfaces have recently been shown as a very promising platform in the field of nonclassical imaging. It includes both the use of metasurfaces for classical imaging of single-photon emitters and imaging with non-classical light.

Huang et al. [Huang2019] developed a quantum platform based on diamond with NV centres, which are known to harbour electron spins that can be manipulated at room temperature, see **Fig. 5e**. Each NV centre emits light that provides information about the spin's quantum state. A metasurface was fabricated on the surface of diamond, acting like metalens to collect photons from a single qubit in diamond and direct them into an optical fibre. This is the first key step in larger efforts to realize compact quantum devices operating at room temperature.

Quantum entanglement is a key resource that can be exploited for a range of applications such as quantum teleportation, quantum computation, and quantum cryptography. The efforts to exploit entanglement in imaging systems have so far led to solutions such as quantum imaging with undetected photons [Lemos2014] and ghost imaging [Pittman1995, Gatti2008, Bornman2019]. The latter however has found useful classical implementations, though with quantum light one would expect better signal-to-noise ratio. Altuzarra et al. [Altuzarra2019] demonstrated an optical imaging protocol that relies uniquely on entanglement. As shown in **Fig. 5f,** two polarizing patterns imprinted and superimposed on a metasurface are separately imaged only when using entangled photons. Unentangled light is not able to distinguish between the two patterns. Entangled photon imaging of functional metasurfaces promises advances towards the use of nanostructured subwavelength devices in quantum information protocols and a route to efficient quantum state tomography.

**Concluding Remarks and Outlook**

The recent strong efforts in developing the field of metasurfaces are being focused on exploring the new physics that can lead to breakthrough applications in quantum photonics. Now in particular, active metasurfaces with external control of their characteristics, and the use of metasurfaces to control quantum light and quantum properties such as single-photon emission and nonclassical detection becomes crucially important.

Metasurfaces integrated with quantum emitters could be used as a special meta-platform for quantum photon sources. In the regime of weak coupling, Purcell enhancement of quantum emitters, or their enhancement of spontaneous emission rates, along with spatial multiplexing and directionality control, could enable the development of efficient single-photon sources for quantum optics. Besides this could also become a sensor of single atoms or molecules. Metasurfaces can be used to manipulate many-body cooperative interactions among the emitters. The strong coupling and ultrastrong coupling of emitters to metasurfaces need to be investigated. We also observe the development of the fundamental and practical advances in the realization of multi-photon quantum interference that takes place at the subwavelength scale. These approaches pave the road for novel types of ultrathin metadevices [Zheludev2012] for the manipulation and measurement of multi-photon quantum-entangled photon states. In addition, the coherent perfect absorption can become a significant technique for monitoring very weak absorption as this would disturb the conditions for perfect absorption and any leaking light would be a measure of such absorption.

Unique properties of metasurfaces to manipulate photons, for example the arrival of two photons together at a port or arrival of one and to produce states of nonclassical light, could find use in a variety of areas, including free-space communications and quantum imaging. A key strength of this system is that it enables complete quantum-state measurements using simple polarization-insensitive single-photon click detectors. Combining the metasurface with single-photon-sensitive charge-coupled device cameras could allow multiple-time-frame images of quantum states. This type of metasurface is thus analogous to "a quantum camera lens" that allows fast imaging-based measurements of quantum states.

Metasurfaces themselves might become a novel type of enabling devices for routing and manipulating nonclassical light in quantum communications, quantum information processing, and quantum computing, thereby paving the way to numerous practical applications including, among others, the development of unbreakable encryption, as well as opening the door to new possibilities for quantum information systems on a chip.

**Acknowledgements**
The authors thank M. Chekhova, M. Davis, J. Ruostekoski, D.P. Tsai, and V. Zadkov for useful comments and suggestions. The authors (ASS and YSK) acknowledge a support from the Australian Research Council (DE180100070 and DP200101168), the University of Technology Sydney (Seed Funding Grant), and the Strategic Fund of the Australian National University. GSA acknowledges support from R. A. Welch Foundation (grant A-1943).

**References**
[Abouraddy2004] A.F. Abouraddy, P.R. Stone, A.V. Sergienko, B.E.A. Saleh, and M.C. Teich, Entangled-photon imaging of a ure phase object, *Phys. Rev. Lett.* **93**, 213903 (2004).

[Agarwal1975] G.S. Agarwal, Quantum electrodynamics in presence of dielectrics and conductors-IV General theory of spontaneous emission in finite geometries, *Phys. Rev. A* **12**, 1474, (1975).

[Agarwal2000] G. S. Agarwal, Anisotropic vacuum-induced interference in decay channels, *Phys. Rev. Lett.* **84**, 5500 (2000).

[Aharonov1988] Y. Aharonov, D. Z. Albert, and L. Vaidman, How the result of a measurement of a component of the spin of a spin-1/2 particle can turn out to be 100, *Phys. Rev. Lett.* **60**, 1351 (1988).

[Altewischer2002] E. Altewischer, M. P. van Exter, and J. P. Woerdman, Plasmon-assisted transmission of entangled photons, *Nature* **14**, 304 (2002).

[Altuzarra2019] C. Altuzarra, A. Lyons, G. Yuan, C. Simpson, T. Roger, J. S. Ben-Benjamin, and D. Faccio, Imaging of polarization-sensitive metasurfaces with quantum entanglement, *Phys. Rev A* **99,** 020101(R) (2019).

[Anderson2017] B. E. Anderson, P. Gupta, B. L. Schmittberger, T. Horrom, C. Hermann-Avigliano, K. M. Jones, and P. D. Lett, Optimized phase sensing in a truncated SU(1,1) interferometer, *Optica* **4**, 752 (2017).

[Ballantine2020] K. E. Ballantine and J. Ruostekoski, Optical magnetism and Huygens' surfaces in arrays of atoms induced by cooperative responses, arXiv: 2002.12930 (2020).

[Baranov2017] D. G. Baranov, A. Krasnok, T. Shegai, A. Alù, and Y. Chong, Coherent perfect absorbers: linear control of light with light, *Nature Reviews Materials* **2**, 17064 (2017).

[Barredo2016] D. Barredo, S. de Léséleuc, V. Lienhard, T. Lahaye, and A. Browaeys, An atom-by-atom assembler of defect-free arbitrary two-dimensional atomic arrays, *Science* **354**, 1021-1023 (2016).

[Bekenstein2020] R. Bekenstein, I. Pikovski, H. Pichler, E. Shahmoon, S. F. Yelin, and M. D. Lukin, Quantum metasurfaces with atom arrays, *Nature Physics* **16**, 676–681 (2020).

[Berraquero2017] C. Palacios-Berraquero, D. M. Kara, A. R.-P. Montblanch, M. Barbone, P. Latawiec, D. Yoon, A. K. Ott, M. Loncar, A. C. Ferrari, and M. Atatüre, Large-scale quantum-emitter arrays in atomically thin semiconductors, *Nature Communications* **8**, 15093 (2017).

[Beugnon2006] J. Beugnon, M. P. A. Jones, J. Dingjan, B. Darquié, G. Messin, A. Browaeys, and P. Grangier, Quantum interference between two single photons emitted by independently trapped atoms, *Nature* **440**, 779–782 (2006).

[Biehs2017] A. Biehs and G. S. Agarwal, Qubit entanglement across near-zero media, *Phys. Rev. A* **96**, 022308 (2017).

[Bornman2019] N. Bornman, M. Agnew, F. Zhu, A. Vallés, A. Forbes, and J. Leach, Ghost imaging using entanglement-swapped photons, *NPJ Quantum Information* **5**, 63 (2019).

[Brida2010] G. Brida, M. Genovese, and I. Ruo Berchera, Experimental realization of sub-shot-noise quantum imaging, *Nature Photonics* **4**, 227–230 (2010).

[Caspani2017] L. Caspani, C. Xiong, B. J. Eggleton, D. Bajoni, M. Liscidini, M. Galli, R. Morandotti, and D.J. Moss, Integrated sources of photon quantum states based on nonlinear optics, *Light: Science & Applications* **6**, 17100 (2017).

[Chang2018] S. Chang, X. Guo, and X. Ni, Optical Metasurfaces: Progress and Applications, *Annual Review of Materials Research* **48**, 279-302 (2018).

[Chen2016] H.-T. Chen, A. J. Taylor, and N. Yu, A review of metasurfaces: physics and applications, *Reports on Progress in Physics* **79**, 76401 (2016).

[Chen2017] S. Chen, X. Zhou, C. Mi, Z. Liu, H. Luo, and S. Wen, Dielectric metasurfaces for quantum weak measurements, *Applied Physics Letters* **110**, 161115 (2017).

[Chen2018] S. Chen, Z. Li, Y. Zhang, H. Cheng, and J. Tian, Phase manipulation of electromagnetic waves with metasurfaces and its applications in nanophotonics, *Advanced Optical Materials* **6**, 1800104 (2018).

[Collett1987] M.J. Collett, R. Loudon, and C.W. Gardiner, Quantum theory of optical homodyne and heterodyne detection, *Journal of Modern Optics*, **34**, 881-902 (1987).

[Dowran2018] M. Dowran, A. Kumar, B. J. Lawrie, R. C. Pooser, and A. M. Marino, Quantum-enhanced plasmonic sensing, *Optica* **5**, 628 (2018).

[Gatti2008] A. Gatti, E. Brambilla, and L. Lugiato, Quantum Imaging, *Prog. In Opt.* **51**, 251, (2008).

[Gefen2019] T. Gefen, A. Rotem, and A. Retzker, Overcoming resolution limits with quantum sensing, *Nature Communications* **10**, 4992 (2019).

[Georgi2019] P. Georgi, M. Massaro, K.-H. Luo, B. Sain, N, Montaut, H. Herrmann, T. Weiss, G. Li, C. Silberhorn, and T. Zentgraf, Metasurface interferometry toward quantum sensors, *Light: Science & Applications* **8**, 70 (2019).

[Giovannetti2004] V. Giovannetti, S. Lloyd, and L. Maccone, Quantum-enhanced measurements: Beating the standard quantum limit, *Science* **306**, 1330 (2004).

[Gisin2002] N. Gisin, G. Ribordy, W. Tittel, and H. Zbinden, Quantum cryptography, *Rev. Mod. Phys.* **74**, 145 (2002).

[Gisin2007] N. Gisin and R. Thew, Quantum communication, *Nature Photonics* **1**, 165–171 (2007).

[Hong1987] C. K. Hong, Z. Y. Ou, and L. Mandel, Measurement of sub-picosecond time intervals between two photons by interference, *Phys Rev Lett.* **59**, 2044 (1987).

[Huang2014] S. Huang and G.S. Agarwal, Coherent perfect absorption of path entangled single photons, *Optics Express* **22**, 20936 (2014).

[Huang2019] T. Y. Huang, R. R. Grote1, S. A. Mann, D. A. Hopper, A. L. Exarhos, G. G. Lopez, G. R. Kaighn, E. C. Garnett, and L. C. Bassett, A monolithic immersion metalens for imaging solid-state quantum emitters, *Nature Communications* **10**, 2392 (2019).

[Hudelist2014] F. Hudelist, J. Kong, C. Liu, J. Jing, Z. Y. Ou, and W. Zhang, Quantum metrology with parametric amplifier-based photon correlation interferometers, *Nature Communications* **5**, 3049 (2014).

[Iwanaga2018] M. Iwanaga, T. Mano, and N. Ikeda, Superlinear photoluminescence dynamics in Plasmon−Quantum- Dot Coupling Systems, *ACS Photonics* **5**, 897−906 (2018).

[Jha2015] P. K. Jha, X. Ni, C. Wu, Y. Wang, and X. Zhang, Metasurface-enabled remote quantum interference, *Phys Rev Lett.* **115**, 025501 (2015).

[Jha2018] P. K. Jha, N. Shitrit, J. Kim, X. Ren, Y. Wang, and X. Zhang, Metasurface-mediated quantum entanglement, *ACS Photonics* **5**, 971−976 (2018).

[Kamali2018] S. Kamali, E. Arbabi, A. Arbabi, and A. Faraon, A review of dielectric optical metasurfaces for wavefront control, *Nanophotonics* **7**, 1041-1068 (2018).

[Kan2020] Y. Kan, S. K. H. Andersen, F. Ding, S. Kumar, C. Zhao, and S. I. Bozhevolnyi, Metasurface-enabled generation of circularly polarized single photons, *Adv. Mater.* **32**, 1907832 (2020).

[Kang2019] L. Kang, R. P. Jenkins, and D. H. Werner, Recent progress in active optical metasurfaces. *Advanced Optical Materials* **7**, 1801813 (2019).

[Kornovan2019] D. Kornovan, M. Petrov, and I. Iorsh, Noninverse dynamics of a quantum emitter coupled to a fully anisotropic environment, *Phys, Rev. A* **100**, 033840 (2019).

[Kruk2017] S. Kruk and Y. Kivshar, Functional meta-optics and nanophotonics governed by Mie resonances, *ACS Photonics* **4**, 2638-2649 (2017).

[Ladd2010] T. Ladd, F. Jelezko, R. Laflamme, Y. Nakamura, C. Monroe, and J. L. O'Brien, Quantum computers. *Nature* **464**, 45–53 (2010).

[Lassalle2020] E. Lassalle, P. Lalanne, S. Aljunid, P. Genevet, B. Stout, T. Durt, and D. Wilkowski, Long-lifetime coherence in a quantum emitter induced by a metasurface, *Phys. Rev. A* **101**, 013837 (2020).

[Leibfried2004] D. Leibfried, M.D. Barrett, T. Schaetz, J. Britton, J. Chiaverini, W.M. Itano, J.D. Jost, C. Langer and D.J. Wineland, Toward Heisenberg-limited spectroscopy with multi-particle entangled states, Science **304**, 1476 (2004).

[Lemos2014] G. B. Lemos, V. Borish, G. D. Cole, S. Ramelow, R. Lapkiewicz, and A. Zeilinger, Quantum imaging with undetected photons, *Nature* **512**, 409 (2014).

[Li_C2020] C. Li, P. Yu, Y. Huang, Q. Zhou, J. Wu, Z. Li, X. Tonga, Q. Wenc, H.-C. Kuod, and Z. M. Wanga, Dielectric metasurfaces: From wavefront shaping to quantum platforms, *Progress in Surface Science* **95**, 100584 (2020).

[Li_G2017] G. Li, S. Zhang, and T. Zentgraf, Nonlinear photonic metasurfaces. *Nature Review Materials* **2**, 17010 (2017).

[Li_L2020] L. Li, Z. Liu, X. Ren, S. Wang, V.-C. Su, M.-K. Chen, C. H. Chu, H. Y. Kuo, B. Liu, W. Zhang, G. Guo, L. Zhang, Z. Wang, S. Zhu, and D.P. Tsai, Metalens-array-based high-dimensional and multiphoton quantum source, *Science* **368**, 1487 (2020).

[Lung2020] S. Lung, K. Wang, K. Z. Kamali, J. Zhang, M. Rahmani, D. N. Neshev, and A. A. Sukhorukov, Complex-birefringent dielectric metasurfaces for arbitrary polarization-pair transformations, arXiv:2006.16559 (2020).

[Lunnemann2016] P. Lunnemann and A. F. Koenderink, The local density of optical states of a metasurface, *Sci. Rep.* **6**, 20655 (2016).

[Luo2018] X. Luo, Subwavelength optical engineering with metasurface waves, *Advanced Optical Materials* **6**, 1701201 (2018).

[Lyons2019] A. Lyons, D. Oren, T. Roger, V. Savinov, J. Valente, S. Vezzoli, N. I. Zheludev, M. Segev, and D. Faccio, Coherent metamaterial absorption of two-photon states with 40% efficiency, *Phys. Rev. A* **99**, 011801(R) (2019).

[Magaña-Loaiza 2019] O.S. Magaña-Loaiza, R. de J. León-Montiel, A. Perez-Leija, A.B. U'Ren, C. You, K. Busch, A.E. Lita, S.W. Nam, R.P. Mirin, and T. Gerrits, Multiphoton quantum-state engineering using conditional measurements, *NPJ Quantum Information* **5**, 80 (2019).

[Marino2019] G. Marino, A. S. Solntsev, L. Xu, V. F. Gili, L. Carletti, A. N. Poddubny, M. Rahmani, D. A. Smirnova, H. Chen, A. Lemaître, G. Zhang, A. V. Zayats, C. De Angelis, G. Leo, A. A. Sukhorukov, and D. N. Neshev, Spontaneous photon-pair generation from a dielectric nanoantenna, *Optica* **6**, 1416-1422 (2019).

[Ming2020] Y. Ming, W. Zhang, J. Tang, Y. Liu, Z. Xia, Y. Liu, and Y.-Q. Lu, Photonic entanglement based on nonlinear metamaterials, *Laser & Photonics Reviews* **14**, 1900146 (2020).

[Mizuochi2012] N. Mizuochi, T. Makino, H. Kato, D. Takeuchi, M. Ogura, H. Okushi, M. Nothaft, P. Neumann, A. Gali, F. Jelezko, J. Wrachtrup, and S. Yamasaki, Electrically driven single-photon source at room temperature in diamond, *Nature Photonics* **6**, 299–303 (2012).

[Moreau2019_2] P. Moreau, E. Toninelli, T. Gregory, and M. J. Padget, Imaging with quantum states of light. *Nat. Rev. Phys.* **1**, 367–380 (2019).

[Moreau2019] P.-A. Moreau, E. Toninelli, T. Gregory, R. S. Aspden, P. A. Morris and M. J. Padgett, Imaging Bell-type nonlocal behaviour, *Science Advances* **5**, eaaw2563 (2019).

[Nagali2009] E. Nagali, F. Sciarrino, F. De Martini, L. Marrucci, B. Piccirillo, E. Karimi, and E. Santamato, Quantum information transfer from spin to orbital angular momentum of photons, *Phys. Rev. Lett*. **103**, 013601 (2009)

[Okoth2019] C. Okoth, A. Cavanna, T. Santiago-Cruz, and M.V. Chekhova, Microscale generation of entangled photons without momentum conservation, *Phys. Rev. Lett.* **123**, 263602 (2019).

[Ono2013] T. Ono, R. Okamoto, and S. Takeuchi, An entanglement-enhanced microscope, Nature Communications **4**, 3426 (2013).

[Osborne2019] I. S. Osborne, Dynamic metasurfaces, *Science* **364**, 645-647 (2019).

[Overvig2019] A. C. Overvig, S. Shrestha, S. C. Malek, M. Lu, A. Stein, C. Zheng, and N. Yu. Dielectric metasurfaces for complete and independent control of the optical amplitude and phase *Light Sci. Appl.* **8**, 92 (2019).

[Pan2012] J.-W. Pan, Z.-B. Chen, C.-Y. Lu, H. Weinfurter, A. Zeilinger, and M. Żukowski, Multiphoton entanglement and interferometry, *Rev. Mod. Phys*. **84**, 777 (2012).

[Parmee2020] C. D. Parmee, J. Ruostekoski, Signatures of optical phase transitions in super- and subradiant arrays of atoms, arXiv:2007.03473 (2020).

[Pelton2015] M. Pelton, Modified spontaneous emission in nanophotonic structures, *Nature Photonics* **9**, 427 (2015).

[Perczel2020] J. Perczel, J. Borregaard, D.E. Chang, S. F. Yelin, and M. D. Lukin, Topological quantum optics using atom-like emitter arrays coupled to photonic crystals, *Physical Review Letters* **124**, 083603 (2020).

[Pittman1995] T. B. Pittman, Y. H. Shih, D. V. Strekalov, and A. V. Sergienko, Optical imaging by means of two-photon quantum entanglement, *Physical Review A* **52**, R3429 (1995).

[Proscia2018] N. V. Proscia, Z. Shotan, H. Jayakumar, P. Reddy, C. Cohen, M. Dollar, A. Alkauskas, M. Doherty, C. A. Meriles, and V. M. Menon, Near-deterministic activation of room-temperature quantum emitters in hexagonal boron nitride, *Optica* **5**, 1128-1134 (2018).

[Pryde2005] G. J. Pryde, J. L. O'Brien, A. G. White, T. C. Ralph, and H. M. Wiseman. Measurement of quantum weak values of photon polarization, *Physical Review Letters* **94**, 220405 (2005).

[Reck1994] M. Reck, A. Zeilinger, H.J. Bernstein, and P. Bertani, Experimental realization of any discrete unitary operator, *Phys. Rev. Lett*. **73**, 58 (1994).

[Ritchie1991] N. W. M. Ritchie, J. G. Story, and R. G. Hulet. Realization of a measurement of a 'weak value'. *Physical Review Letters* **66**, 1107–1110 (1991).

[Roger2015] T. Roger, S. Vezzoli, E. Bolduc, J. Valente, J. J. F. Heitz, J. Jeffers, C. Soci, J. Leach, C. Couteau, N. I. Zheludev, and D. Faccio, Coherent perfect absorption in deeply subwavelength films in the single-photon regime, *Nature Communications* **6**, 7031 (2015).

[Rubin2019] N. A. Rubin, G. D'Aversa, P. Chevalier, Z. Shi, W.T. Chen, and F. Capasso, Matrix Fourier optics enables a compact full-Stokes polarization camera, *Science* **365**, 43 (2019).

[Rui2020] J. Rui, D. Wei, A. Rubio-Abadal, S. Hollerith, J. Zeiher, D. M. Stamper-Kurn, C. Gross, and I. Bloch, A subradiant optical mirror formed by a single structured atomic layer, *Nature* **583**, 369–374 (2020).

[Salvail2013] J. Z. Salvail, M. Agnew, A. S. Johnson, E. Bolduc, J. Leach, and R. W. Boyd. Full characterization of polarization states of light via direct measurement, *Nature Photonics* **7**, 316–321 (2013).

[Samantaray2017] N. Samantaray, I. Ruo-Berchera, A. Meda, and M. Genovese, Realization of the first sub-shot-noise wide field microscope, *Light: Science & Applications* **6**, e17005 (2017).

[Shields2007] A. J. Shields, Semiconductor quantum light sources, *Nature Photonics* **1**, 215 (2007).

[Slussarenko2019] S. Slussarenko and G. J. Pryde, Photonic quantum information processing: A concise review, *Applied Physics Reviews* **6**, 041303 (2019).

[Solntsev2017] A. S. Solntsev and A. A. Sukhorukov, Path-entangled photon sources on nonlinear chips, *Reviews in Physics* **2**, 19-31 (2017).

[Solntsev2018] A. S. Solntsev, P. Kumar, T. Pertsch, A. A. Sukhorukov, and F. Setzpfandt, LiNbO3 waveguides for integrated SPDC spectroscopy, *APL Photonics* **3**, 021301 (2018).

[Stav2018] T. Stav, A. Faerman, E. Maguid, D. Oren, V. Kleiner, E. Hasman, and M. Segev, Quantum entanglement of the spin and orbital angular momentum of photons using metamaterials, *Science* **361**, 1101–1104 (2018).

[Strekalov1995] D. V. Strekalov, A. V. Sergienko, D. N. Klyshko, and Y. H. Shih, Observation of two-photon "ghost" interference and diffraction, *Phys. Rev. Lett.* **74**, 3600 (1995).

[Suchowski2013] H. Suchowski, K. O'Brien, Z.J. Wong, A. Salandrino, X. Yin, and X. Zhang, Phase mismatch–free nonlinear propagation in optical zero-index materials, *Science* **342**, 1223 (2013).

[Tran2016] T. T. Tran, K. Bray, M. J. Ford, M. Toth, and I. Aharonovich, Quantum emission from hexagonal boron nitride monolayers, *Nature Nanotechnology* **11**, 37–41 (2016).

[Tran2017] T. T. Tran, D. Wang, Z.-Q. Xu, A. Yang, M. Toth, T. W. Odom, and I. Aharonovich, Deterministic coupling of quantum emitters in 2D materials to plasmonic nanocavity arrays, *Nano Lett.* **17**, 2634−2639 (2017).

[Varro2008] S. Varro, Correlations in single-photon experiments, *Fortschr. Phys*. **56**, 91-102 (2008).

[Vaskin2019] A. Vaskin, R. Kolkowskia, A. F. Koenderink, and I. Staude, Light-emitting metasurfaces, *Nanophotonics* **8**, 1151–1198 (2019).

[Wang2018] K. Wang, J. G. Titchener, S. S. Kruk, L. Xu, H.-P. Chung, M. Parry, I. I. Kravchenko, Y.-H. Chen, A. S. Solntsev, Y. S. Kivshar, D. N. Neshev, and A. A. Sukhorukov, Quantum metasurface for multiphoton interference and state reconstruction, *Science* **361**, 1104–1108 (2018).

[Wen2013] J. Wen, Y. Zhang, and M. Xiao, The Talbot effect: recent advances in classical optics, nonlinear optics, and quantum optics, *Advances in Optics and Photonics* **5**, 83-130 (2013).

[Xie2020] Y.-Y. Xie, P.-N. Ni, Q.-H. Wang, Q. Kan, G. Briere, P.-P. Chen, Z.-Z. Zhao, A. Delga, H.-R. Ren, H.-D. Chen, C. Xu, and P. Genevet, Metasurface-integrated vertical cavity surface-emitting lasers for programmable directional lasing emissions, *Nature Nanotechnology* **15**, 125–130 (2020).

[Xu2018] D. Xu, X. Xiong, L. Wu, X.-F. Ren, C. E. Peng, G.-C. Guo, Q. Gong, and Y.-F. Xiao, Quantum plasmonics: new opportunity in fundamental and applied photonics, *Advances in Optics and Photonics* **10**, 703-756 (2018).

[Yin2017] J. Yin, Y. Cao, Y.-H. Li, et al., Satellite-based entanglement distribution over 1200 kilometres, *Science* **356**, Issue 1140–1144 (2017).

[Yin2020] J. Yin, Y. Li, S. Liao, et al., Entanglement-based secure quantum cryptography over 1,120 kilometres. *Nature* **582**, 501–505 (2020).

[Yu 2014] N. Yu and F. Capasso, Flat optics with designer metasurfaces, *Nature Materials* **13**, 139 (2014).

[Zheludev2012] N. I. Zheludev and Y. S. Kivshar, From metamaterials to metadevices, *Nature Materials* **11**, 917-924 (2012).

[Zhu2020] L. Zhu, X. Liu, B. Sain, M. Wang, C. Schlickriede, Y. Tang, J. Deng, K. Li, J. Yang, M. Holynski, S. Zhang, T. Zentgraf, K. Bongs, Y.-H. Lien, and G. Li, A Dielectric Metasurface Optical Chip for the Generation of Cold Atoms, *Science Advances* **6**, accepted (2020).